\documentclass[12pt,reprint,aps,groupedaddress,nofootinbib,prd,twocolumn]{revtex4-1}

\usepackage[english]{babel}
\usepackage[utf8]{inputenc}
\usepackage{amsmath}
\usepackage{mathbbol}
\usepackage{amssymb}
\usepackage{bbold}
\usepackage{graphicx,amsfonts}
\usepackage{epsfig}
\usepackage{xcolor}
\definecolor{crimsonglory}{rgb}{0.75,0.0,0.2}
\usepackage[colorlinks=true,
linkcolor=crimsonglory,
urlcolor=crimsonglory,
citecolor=crimsonglory]{hyperref}
\usepackage{capt-of}
\usepackage{bm}
\usepackage{mathrsfs}
\usepackage{enumerate}
\usepackage{amsthm}
\usepackage{bbm}
\usepackage{comment}
\usepackage{appendix}
\usepackage{physics}
\usepackage{url}
\usepackage{upgreek}

\def\rotatechartwo#1{\reflectbox{#1}}
\newcommand{\be}{\begin{equation}}
	\newcommand{\ee}{\end{equation}}
\newcommand{\beq}{\begin{equation}}
	\newcommand{\eeq}{\end{equation}}
\newcommand{\bea}{\begin{eqnarray}}
	\newcommand{\eea}{\end{eqnarray}}
\newcommand{\bit}{\begin{itemize}}
	\newcommand{\eit}{\end{itemize}}
\newcommand{\ben}{\begin{enumerate}}
	\newcommand{\een}{\end{enumerate}}

\newcommand{\derr}[2]{\frac{\mathrm{d}^2 #1}{\mathrm{d} #2 ^2}}

\newcommand{\de}[1]{\mathrm{d} #1}

\begin{document}
\title{\textcolor{red}{RE\rotatechartwo{D}shift}: quasinormal modes of black holes embedded in halos of matter}
	\author{Laura Pezzella$^{1,2,3,4}$, Kyriakos Destounis$^{1,2,5}$, Andrea Maselli$^{3,4}$, Vitor Cardoso$^{5,6}$}
	\affiliation{$^1$Dipartimento di Fisica, Sapienza Università di Roma, Piazzale Aldo Moro 5, 00185, Roma, Italy}
	\affiliation{$^2$INFN, Sezione di Roma, Piazzale Aldo Moro 2, 00185, Roma, Italy}
	\affiliation{$^3$Gran Sasso Science Institute (GSSI), I-67100 L’Aquila, Italy}
	\affiliation{$^4$INFN, Laboratori Nazionali del Gran Sasso, I-67100 Assergi, Italy}
	\affiliation{$^5$CENTRA, Departamento de F\'isica, Instituto Superior T\'ecnico – IST, Universidade de Lisboa – UL, Avenida Rovisco Pais 1, 1049-001 Lisboa, Portugal}
	\affiliation{$^6$Niels Bohr International Academy, Niels Bohr Institute, Blegdamsvej 17, 2100 Copenhagen, Denmark}
	
	\begin{abstract}
		We investigate the (axial) quasinormal modes of black holes embedded in generic matter profiles. Our results reveal that the axial QNMs experience a redshift when the black hole is surrounded by various matter environments, proportional to the compactness of the matter halo. Our calculations demonstrate that for static black holes embedded in galactic matter distributions, there exists a universal relation between the matter environment and the redshifted vacuum quasinormal modes. In particular, for dilute environments the leading order effect is a redshift $1+U$ of frequencies and damping times, with $U \sim -{\cal C}$ the Newtonian potential of the environment at its center, which scales with its compactness ${\cal C}$. 
	\end{abstract}
	
	\maketitle
	
	\section{Introduction}
	For decades, black holes (BHs) have only been considered as mathematical solutions to the Einstein field equations, without any astrophysical relevance. It took many years before they were accepted as anything other than a mathematical curiosity. Even though \emph{``there should be a law of nature to prevent a star from behaving in this absurd way''}\footnote{Arthur Eddington, Royal Astronomical Society conference, January 1935}, we now are certain that BHs not only exist but also directly affect the way we understand the cosmos in its most extreme regimes. Recent experiments detected gravitational waves (GWs) from binary BHs and neutron stars \cite{KAGRA:2021vkt,LIGO:2021ppb}. These detections are now characterized as the pinnacle of GW astronomy \cite{Schutz:1999xj,Bishop:2021rye,Bailes:2021tot}. Combined with the foundational results of Gravity Probe A, B \cite{Vessot:1980zz,Everitt:2011hp}, astrometry measurements at the center of our galaxy \cite{Ghez:1998ph,Genzel:2010}, X-ray spectroscopy \cite{Canizares:1979,Canizares:1982,Baganoff:2003}, infrared spectroscopy \cite{Boughn:1983,Perlman:2007zc} and the groundbreaking measurement of the shadows of supermassive BHs in M87* and Sgr A* with the Event Horizon Telescope \cite{EventHorizonTelescope:2019dse,EventHorizonTelescope:2022wkp}, the direct detection of GWs established the field of multimessenger astrophysics \cite{Meszaros:2019xej}. BHs do not only exist in nature, but also interact with other celestial objects, surrounding matter and the interstellar medium that veils the universe. 
	
	Decades of simulations and observations have shown that any event involving a binary BH merger is likely to end in the same way; the GW amplitude of the merger will decay as a superposition of characteristic damped sinusoids. The damped oscillations are described by quasinormal modes (QNMs) \cite{Kokkotas:1999bd,Berti:2009kk,Konoplya:2011qq}, i.e. a complex set of frequencies which are an intrinsic property of dissipative systems. Calculating QNMs is achieved with the use of BH perturbation theory, and has led to the establishment of the BH spectroscopy program \cite{Berti:2005ys,Giesler:2019uxc,Baibhav:2023clw,Destounis:2023ruj,Ringdown_review}. QNM frequencies are eigenmodes that are linked to the generic response of a BH under small fluctuations to the background spacetime and can help us to understand the nature of the remnant, providing 
	precise measurements of its mass and spin, as well as allowing for tests of General Relativity (GR) \cite{Berti:2005ys,Berti:2009kk,Cardoso:2019rvt,Franchini:2023eda}. 
	
	The striking majority of QNM literature analyzes the spectral content of \emph{vacuum} BHs, clearly an idealization. Massive BHs sit at the center of galaxies which are orders of magnitude more massive, where gaseous accretion disks \cite{Duschl:1989,Murchikova:2019,Speri:2022upm}, dense dark matter spikes \cite{Daghigh:2022pcr,Chan:2024}, and galactic dark matter halos form that affect the whole galaxy \cite{Wechsler:2018pic}. When two galaxies merge, baryonic and dark matter will interact and the supermassive BHs in their cores will meld into a new supermassive BH~\cite{Struck:1999mf}; this event is expected to emit mHz GWs and falls into the frequency band of spaceborne interferometers, like the Laser Interferometer Space Antenna (LISA)~\cite{LISA:2017pwj,LISA:2022yao,LISA:2022kgy,Karnesis:2022vdp}. In this extreme case, the entire galaxy is an active participant in the dynamics and relaxation of the BH(s). The universe is also permeated with interstellar plasma, which prevents propagation of low-frequency electromagnetic waves, a screening mechanism occurs that entirely cloaks electromagnetic from GWs, which can have strong implications in BH spectroscopy \cite{Cannizzaro:2024yee}. 
	
	Taking into account astrophysical environments around BHs leads to interesting effects that can affect the emission and propagation of GWs. Recently, an interesting phenomenon -- \emph{spectral instability} -- regained a lot of interest and is directly tied to the existence of astrophysical environments around BHs. Spectral instabilities occur when QNM excessively migrate in the complex plane. This happens when environmental fluctuations, such as thin matter shells or dark matter spikes, are introduced in the vicinity of vacuum BHs. It first appeared in Refs. \cite{Nollert:1996rf,Nollert:1998ys}, and eventually led to the emergence of an extensive literature in relation to spectral instabilities \cite{Jaramillo:2020tuu,Cheung:2021bol,Berti:2022xfj,Courty:2023rxk,Rosato:2024arw,Oshita:2024fzf} and the use of non-modal tools to complement BH spectroscopy and QNMs, such as the pseudospectrum \cite{Jaramillo:2021tmt,Destounis:2021lum,Gasperin:2021kfv,Jaramillo:2022kuv,Boyanov:2022ark,Destounis:2023ruj,Destounis:2023nmb,Boyanov:2023qqf,Cardoso:2024mrw,Arean:2023ejh,Cownden:2023dam,Sarkar:2023rhp,Cao:2024oud,Luo:2024dxl}. Thus, treating compact objects in non-vacuum systems will provide us with information regarding astrophysical environments and their phenomenology in GW astronomy \cite{Barausse:2014tra}. 
	
	In this work, we aim in giving an extended analysis of axial QNMs for numerically-constructed BH spacetimes with different matter profiles, such as the Hernquist \cite{Hernquist:1990} and Navarro-Frenk-White (NFW) distribution \cite{Navarro:1995iw} that are suited for different kinds of environments. Our expectation is first to provide a public code that can treat axial QNM calculations when any form of matter distribution is present, and to inspect if the redshift effect found for BHs surrounded by Hernquist-type matter \cite{Ringdown_review}, when the compactness is varied, occurs for other matter profiles, thus extending the results regarding the detectability of astrophysical environments with axial QNMs \cite{Spieksma:2024voy}. In what follows, we use geometrized units such that $G=c=1$.
	
	\section{BHs in generic matter halos}
	
	Surrounding matter around perturbed astrophysical BHs can affect their GW emission and propagation. The research on non-vacuum BH QNMs \cite{Leung:1997was,Leung:1999iq} is still in full bloom, with particular focus on galactic supermassive BHs \cite{Cardoso:2021wlq,Konoplya:2021ube}. Most recently, an exact general-relativistic solution of the field equations was found that describes a Schwarzschild BH embedded in a matter halo, that poses the most massive part of any galaxy. Although this approach sounds simplistic, since it does not account for luminous matter and accretion disks, to date only one exact solution of such a static spacetime has been found, where the matter halo is described by the energy density of the Hernquist profile \cite{Hernquist:1990,Cardoso:2021wlq}. In general, more spacetimes of the same nature have been constructed but only numerically or with various approximate techniques \cite{Jusufi:2022jxu,Konoplya:2022hbl,Feng:2022evy,Figueiredo:2023gas,Stelea:2023yqo,Heydari-Fard:2024wgu,Speeney:2024mas,Datta:2023zmd,Shen:2023erj,Maeda:2024tsg}. These fully general-relativistic configurations can, overall, provide a good description of the geometry of galaxies harboring supermassive BHs and have been proven very useful to constrain the environmental surrounding of BHs \cite{Cardoso:2021wlq}. In what follows we describe them briefly.
	
	\subsection{Hernquist-type matter around 
		static BHs}\label{sec:DManalytic}
	
	An exact, fully-relativistic solution that describes a BH immersed in a halo of matter was obtained recently~\cite{Cardoso:2021wlq}. The environmental contribution in this case is accounted for through the Einstein cluster approach, which describes the matter content in terms of an anisotropic stress-energy tensor, with only tangential ($P_t$) and vanishing radial pressure ($P_r=0$), i.e. $T^{\mu}{_{\nu}} = \text{diag}(-\rho,0,P_t,P_t)$. Choosing the energy density of the fluid to follow the Hernquist distribution \cite{Hernquist:1990}, i.e.,
	\begin{equation}
		\rho= \frac{M (a_0+2M_{\rm BH})(1-2M_{\rm BH}/r)}{2 \pi r (r+a_0)^3},\label{Hernquist_profile}
	\end{equation}
	where $M$ and $a_0$ are the halo mass and length scale, allows one to obtain a BH solution in a closed \emph{analytical} form. Observations and galaxy formation simulations indicate that Eq.~\eqref{Hernquist_profile} could be a good proxy for elliptical galaxies and galactic bulges. The geometry of spacetime is given by the metric 
	\begin{equation}\label{eq:uncharged_metric}
		ds^2 = - f(r) dt^2 + \frac{dr^2}{1-\frac{2m(r)}{r}} + r^2(d \theta^2 + \sin^2 \theta d \varphi^2),
	\end{equation}
	where the mass function $m(r)$ is of the form
	\begin{equation}\label{mass_function_m(r)}
		m(r) = M_{\textrm{BH}} + \frac{Mr^2}{(a_0 + r)^2}\Bigg( 1 - \frac{2 M_{\textrm{BH}}}{r} \Bigg)^2,
	\end{equation}
	where $M_{\textrm{BH}}$ is the BH mass. The factor $1-2M_{\rm BH}/r$ in the equations above is chosen to truncate the matter at the BH horizon. It can be chosen freely, in Eq.~\eqref{Hernquist_profile} at the expense of not having a \emph{closed-form} solution for the spacetime geometry. Other truncation factors, such as $1-4M_{\rm BH}/r$ will mimic matter distributions which conform better with relativistic evolutions of dark matter profiles~\cite{Gondolo:1999,Sadeghian:2013}.
	
	At small distances, Eq.~\eqref{mass_function_m(r)} 
	describes a source of mass $M_{\textrm{BH}}$, while at large distances it is a spacetime of ADM mass $M_\textrm{ADM}=M+M_{\rm BH}$. The lapse function $f(r)$ can be obtained from $m(r)$ by imposing that the $(r,r)$ component of Einstein's equations is zero:
	\begin{equation}\label{eq:f_Cardoso}
		f(r) = \Bigg( 1 - \frac{2 M_{\textrm{BH}}}{r} \Bigg) e^{\Upsilon},
	\end{equation}
	with 
	\begin{equation}
		\begin{split}
			\Upsilon &= - \pi \sqrt{\frac{M}{\xi}} + 2 \sqrt{\frac{M}{\xi}}\arctan \frac{r + a_0 - M}{\sqrt{M \xi}},\\
			\xi &= 2 a_0 - M + 4M_{\textrm{BH}}.
		\end{split}    
	\end{equation}
	The BH spacetime has an event horizon at $r = r_h =2M_{\textrm{BH}}$ and a curvature singularity at $r=0$. Assuming $\xi>0$ and a hierarchy of scales, i.e. $M_{\textrm{BH}} \ll M \ll a_0$, ensures that there are no curvature singularities outside the horizon. In the context of dark matter physics, we note that the configuration has a peak tangential pressure $P\lesssim M \rho/a_0$ away from the BH, thus it is Newtonian and may mimic cold dark matter well.
	
	Equation \eqref{eq:f_Cardoso} highlights a generic feature of the geometry of BHs surrounded by {\it any} spherically symmetric matter distribution. In the weak-field limit, this geometry satisfies
	\be
	f(r)=\left(1-\frac{2M_{\rm BH}}{r}\right)e^{2U}\label{weak_field_geometry},
	\ee
	with $U$ being the Newtonian potential of the matter distribution \cite{Duque:2023seg}. For the Hernquist distribution discussed here, $\Upsilon=2U$. We will use Eq. \eqref{weak_field_geometry} to calculate the corrections to the light ring and its frequency, and determine the QNM frequencies in the eikonal regime.
	
	\subsection{Static BHs surrounded by generic matter profiles: a numerical study}\label{numerical_method}
	The Hernquist distribution \eqref{Hernquist_profile} belongs to a more general class of two-parameter profiles, described by the following semi-analytic 
	relation:
	\begin{equation}\label{eq:density_distr}
		\rho(r)=\rho_0 (r/a_0)^{-\gamma}\left[1+\left(r/a_0\right)^\alpha\right]^{(\gamma-\beta)/\alpha} \ .
	\end{equation}
	where $\rho_0$ is the energy density at $r=a_0$. 
	The exponents $\beta$ and $\gamma$ control the slope of the distribution at large and small distances, respectively, while $\alpha$ determines the sharpness of the slope transition \cite{Merritt:2006}. Equation \eqref{eq:density_distr} motivated Refs.~\cite{Figueiredo:2023gas,Speeney:2024mas} 
	to develop a numerical pipeline to constructs static, and spherically symmetric BH solutions embedded in spherically symmetric, but otherwise generic, astrophysical environments. 
	
	In this work, we employ the numerical approach developed in \cite{Figueiredo:2023gas,Speeney:2024mas} and analyze two specific combinations of the $(\alpha, \beta, \gamma)$ parameters from Eq.~\eqref{eq:density_distr} in order to build numerical BH solutions: (i) $(\alpha, \beta, \gamma) = (1, 4, 1)$, which represents the Hernquist distribution, i.e. the numerical ``Hernquist-type BHs'', and (ii) $(\alpha, \beta, \gamma) = (1, 3, 1)$, corresponding to the numerical ``NFW-type BHs'' where the distribution of matter follows the NFW profile. Since the total mass of the NFW profile is logarithmically divergent, we define a radial cutoff $r_c$, such that $M(r>r_c)=0$~\cite{Figueiredo:2023gas}. Hereafter we fix $r_c=5 a_0$. Finally, to mimic the overdense cusp expected to form due to adiabatic accretion, and truncate the matter distribution at the horizon, we scale the profiles such that $\rho(r)\rightarrow \left(1-2M_\mathrm{BH}/r\right)\rho(r)$ \cite{Cardoso:2021wlq,Figueiredo:2023gas,Speeney:2024mas}.
	
	\begin{figure}[t]
		\includegraphics[width=0.5\textwidth]{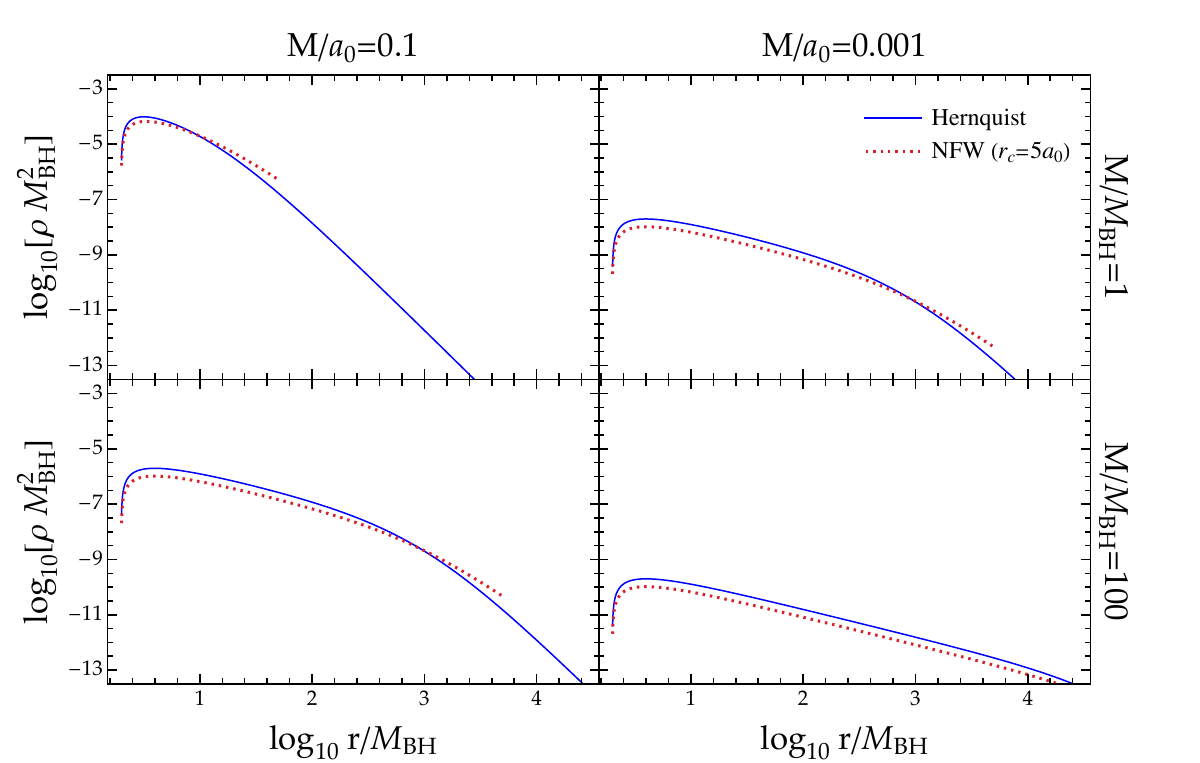}
		\caption{Energy density profiles of numerically-constructed BHs with matter halos, with varying compactness (left and right columns) and varying halo masses (top and bottom rows). The profiles are described by the analytic relation \eqref{eq:density_distr}. We show with blue the energy density of the Hernquist profile [where $(\alpha,\beta,\gamma)=(1,4,1)$] and with dashed red the energy density of the NFW profile [where $(\alpha,\beta,\gamma)=(1,3,1)$], with radial cutoff at $r_c=5a_0$.}\label{fig:energy_density}
	\end{figure}
	
	Figure \ref{fig:energy_density} shows the density distribution $\rho(r)$ for the Hernquist and NFW models, when varying the compactness $M/a_0$ and the mass of the halo.
	
	The BH solutions embedded within the profiles 
	\eqref{eq:density_distr} are constructed as follows:
	\begin{enumerate}
		\item Take the spherically-symmetric, static spacetime described by line element in Eq. \eqref{eq:uncharged_metric}, with $m(r)$ a generic function. 
		\item Prescribe the density profile $\rho(r)$. The mass profile is recovered by solving  the continuity equation $m'(r)=4\pi r^2 \rho(r)$.
		\item The metric function $f(r)$ and the tangential pressure $P_t(r)$ are determined from the $G_{rr}$ component from the Bianchi identities, 
		respectively, as
		\begin{equation}
			\label{eq:f_and_Pt}
			\frac{f'(r)}{f(r)}=\frac{2m(r)/r}{r-2m(r)}, \quad 2P_t(r)=\frac{m(r)\rho(r)}{r-2m(r)}. 
		\end{equation}
	\end{enumerate}
	For a given $\rho(r)$ the solutions to the continuity equation and of Eqs.~\eqref{eq:f_and_Pt} fully specify the background spacetime. 
	
	The ADM mass of the spacetime is given by 
	$M_\textrm{ADM}=M+M_\textup{BH}=m(\infty)$, where $M$ is the total mass contained within the halo. At the horizon the metric approaches the Schwarzschild geometry, with vanishing matter density.
	
	\subsection{Axial perturbations}
\begin{figure}[t]
		\includegraphics[width=0.5\textwidth]{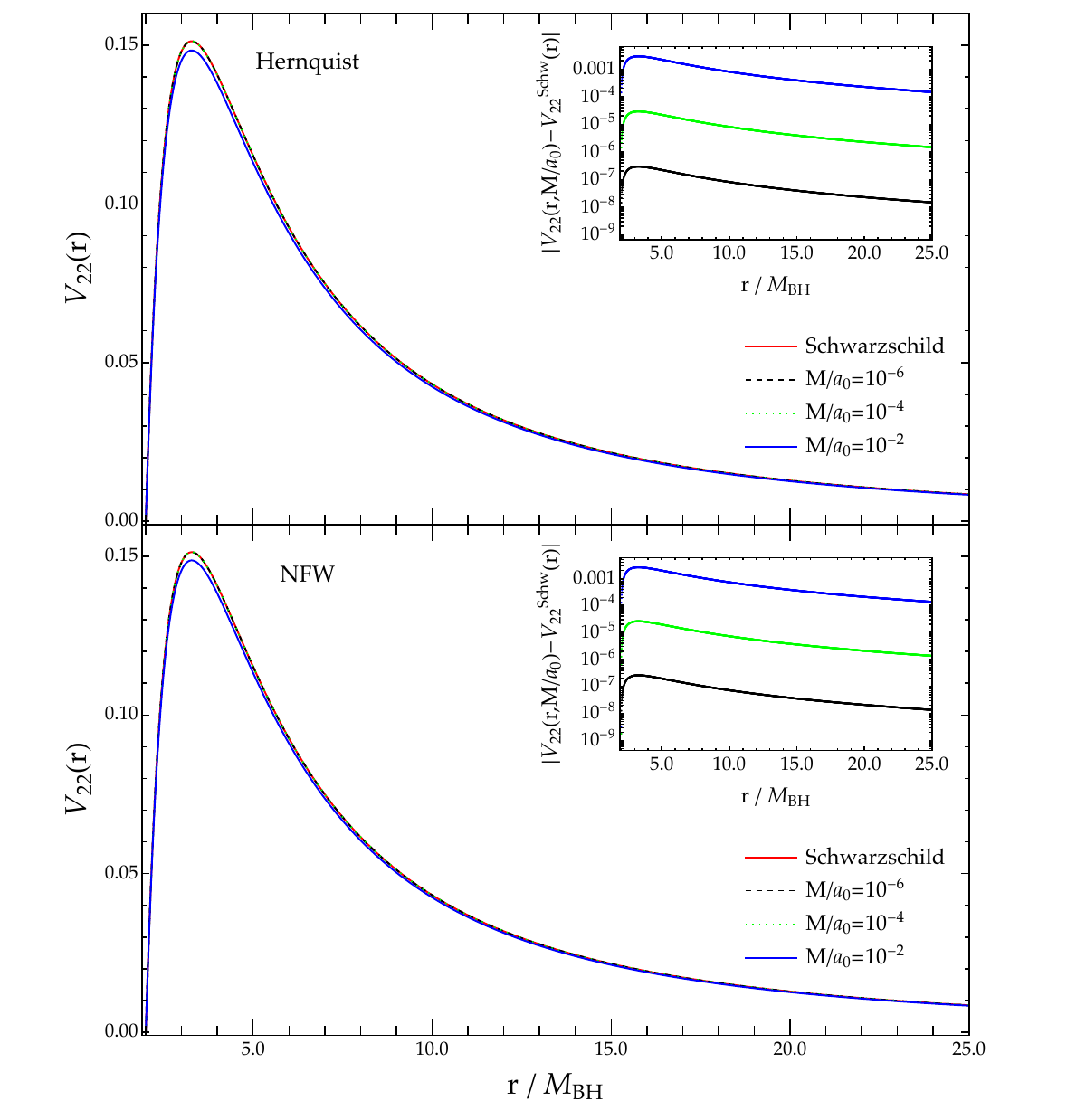}
		\caption{Scattering potential for spin $s=\pm2$ and $\ell=2$ for a Hernquist-type BH (upper panel) and a NFW-type BH (bottom panel) as a function of the radial coordinate $r$ normalized by the black hole mass. The absolute difference with respect to the Schwarzshild case is also shown. The compactness shown are $M/a_0=(10^{-6}, 10^{-4}, \times 10^{-2})$. 
		}\label{fig:Potential}
	\end{figure}

	With a spacetime in hand, we can assess the changes in the QNM spectrum of BHs due to environmental effects. For this, we need to treat linearized fluctuations of the spacetimes we just discussed. Consider perturbations of the metric and stress-energy tensors:
	\begin{equation}
		g_{\mu\nu}=g_{\mu\nu}^{(0)}+g_{\mu\nu}^{(1)} \quad,\quad T_{\mu\nu}^\mathrm{env}=T_{\mu\nu}^{(0) \mathrm{env}}+T_{\mu\nu}^{(1)\mathrm{env}}.
		\label{eq:perturbations}
	\end{equation}
	The first-order components are decomposed in terms of standard axial and polar tensor spherical harmonics \cite{Regge:1957td,Zerilli:1970,Lindblom:1983} that satisfy the perturbed field equations
	\begin{equation}
		G_{\mu\nu}^{(1)}=8\pi T_{\mu\nu}^{(1)\mathrm{env}}.
	\end{equation}
	In this work we focus on axial type perturbations, for which metric and fluid fluctuations decouple (under the assumption that there are no dissipation mechanisms in the fluid~\cite{Boyanov:2024jge}). Then, axial-type gravitational fluctuations are completely governed by a single, one-dimensional, Schrodinger-like equation for a master function $\psi_{\ell m }(r)$~\cite{Cardoso:2022whc,Figueiredo:2023gas}
	\begin{equation}
		\label{eq:master}
		\derr{\psi_{\ell m }(r)}{r_*} + \left(\omega^2 - V_\ell(r)\right) \psi_{\ell m }(r) = 0,
	\end{equation}
	where $\ell=2,\dots, \infty$, $m= -\ell, \dots, \ell$ and $r_*$ is the tortoise coordinate, defined as $\de{r_*}/\de{r}=\left[f(r) (1-2m(r)/r)\right]^{-1/2}$. The scattering potential $V_\ell(r)$ reads \cite{Cardoso:2022whc,Figueiredo:2023gas}:
	\begin{equation}
		V_\ell(r) = f(r) \left(\frac{\ell(\ell+1)}{r^2}-\frac{6 m(r)}{r^3} + \frac{m'(r)}{r^2} \right).
		\label{eq:Pot}
	\end{equation}
	When $m(r)=M_\mathrm{BH}$ we recover the usual vacuum Schwarzschild background and Eq. (\ref{eq:master}) reduces to the well-known Regge-Wheeler equation \cite{Regge:1957td}. Figure~\ref{fig:Potential} shows the scattering potential as a function of the radial coordinate for the Hernquist and the NFW profile, assuming $\ell=2$ and different values of the halo compactness. The inset in each plot shows the relative difference Eq.~\eqref{eq:Pot} and the vacuum limit.
	
	Although not demonstrated in this paper, we have found that Eq.~\eqref{eq:Pot} represents a specific class of a universal scattering potential that depends on the spin $s$ of the perturbing field:
	\begin{equation}
		\label{eq:GenPot}
		V_{s\ell} = f(r) \left[\frac{\ell(\ell+1)}{r^2}+\frac{2 m(r) (1-s^2)}{r^3} - \frac{m'(r)(1-|s|)}{r^2} \right]\ .
	\end{equation} 
	Equation (\ref{eq:GenPot}) generalizes the well known results for Schwarzschild BHs \cite{Berti:2009kk} with a constant mass term, i.e. $m(r)=M_{\rm BH}$. Scalar and electromagnetic perturbations are described by $s=0$ and $s=\pm 1$, respectively, while for gravitational (axial) modes, $s=\pm2$, we recover Eq.~\eqref{eq:Pot}. 
	
	\section{Setup for QNM calculations}
	
	We compute the QNM frequencies for the numerical profiles described in Sec.~\ref{numerical_method}, using the exact Hernquist model of \ref{sec:DManalytic} as a benchmark to validate our procedure (see Appendix~\ref{App:convergence} for a convergence 
	study).
	
	To integrate Eq.~\eqref{eq:master} we first solve the background equations, according to the following two-step procedure:
	\begin{enumerate}
		\item We choose a density profile $\rho (r)$ according to the Hernquist or NFW model. We integrate the equations for the mass function $m(r)$ from the horizon $r_\textup{h}=2 M_\mathrm{BH}$ where $m(r_\textup{h})=M_\mathrm{BH}$ to a coordinate radius $r_\textup{out} \gtrsim 10^7 a_0$ that corresponds to our spatial infinity and guarantees asymptotic flatness. We then solve backwards the equation for the metric function $f(r)$, imposing that, in the far-field limit, $m(r\rightarrow r_\textup{out})=M_\textrm{ADM}$ 
		and the radial function can be written as
		\begin{equation}
			f(r)\simeq 1-\frac{2 M_\mathrm{ADM}}{r}+\mathcal{O}(r^{-3}).
		\end{equation} 
		\item The numerical solution for $m(r)$ and $a(r)$ allows to compute the tangential pressure $P_t(r)$.
	\end{enumerate}    
	QNM calculations require to impose boundary conditions at the horizon and at the infinity. To improve the numerical integration along our domain, we compactify the radial interval by performing the coordinate transformation $x=1-r_\textup{h}/r$ with $x\in[0,1]$, where $x=0$ corresponds to the BH horizon and $x=1$ to spatial infinity. 
	
	At $x=0$ we assume the solution describes a purely ingoing wave:
	\begin{equation}
		\psi_{\ell m,\textup{h}}(x)=x^{-\frac{i \omega r_\textup{h}}{\sqrt{r_\textup{h} f'(r_\textup{h})}}},\label{eq:BCin}
	\end{equation}
	while purely outgoing boundary condition at spatial infinity is satisfied by the relation
	\begin{equation}
		\psi_{\ell m,\rm out} (x)= e^{\frac{i r_\textup{h} \omega}{1-x}} (1-x)^{- 2 i M_\mathrm{ADM}}.\label{eq:BCout}
	\end{equation}
	The ansatz for the wave function reduces to 
	\begin{equation}
		\psi_{\ell m}(x)\sim e^{\frac{i r_\textup{h} \omega}{1-x}} (1-x)^{- 2 i M_\mathrm{ADM}} x^{-\frac{i \omega r_\textup{h}}{\sqrt{r_\textup{h} f'(r_\textup{h})}}}\ .
		\label{eq:boundNum}
	\end{equation}
	This expression generalizes the boundary condition obtained both for the Schwarzschild BH and the exact Hernquist-type BH \cite{Figueiredo:2023gas}. 
	
	We integrate the master equation (\ref{eq:master}) for $\psi_{\ell m}$ supplied by the boundary conditions \eqref{eq:BCin}-\eqref{eq:BCout} using the generalized matrix method described in the following section.
	
	\subsection{Generalized matrix method for QNMs}
	\label{numerical_matrix}
	Matrix methods are powerful computational schemes that transform one-dimensional master equations like Eq.~\eqref{eq:master} into a matrix equation. The basis of matrix methods is the manipulation of the ordinary differential equation that governs perturbation propagation through proper decomposition of the spatial derivatives. The equation is discretized with $N$ equispaced (or otherwise) grid points and transformed into a matrix equation which can be recast in terms of eigenvalues and eigenvectors. 
	
	The matrix method \cite{Lin:2016,Lin:2016sch,Lin:2017oag,Lin:2019mmf,Lin:2022ynv,Shen:2022ssv,Lin:2023rkd,Shen:2024pak} that is utilized and generalized here makes use of grid points in a small region of a query point to estimate its derivatives by employing a Taylor expansion. A key step of the method is to discretize the unknown eigenfunction in order to transform the differential equation and its boundary conditions into a homogeneous matrix equation. Based on the information about $N$ scattered grid points, Taylor series are carried out for the unknown eigenfunction up to $N$-th order for each grid point. The resulting homogeneous system of linear algebraic equations is solved for the eigenvalues.
	
	Here we generalize, for the first time, the matrix method \cite{Lin:2016,Lin:2016sch} for generic, numerically-constructed and spherically-symmetric BH metrics, as explained in Sec. \ref{numerical_method}. In a nutshell, when the numerical BH solutions are built, the result is a vector for each of the metric tensor components. Each vector is \emph{interpolated} with high-accuracy through the standard interpolation formula, which finds a value between two points on the curve of a function. The resulting interpolated metric tensor components can then be accessed at any point of the interpolation function, thus giving us the ability to be re-discretized in order for the matrix method to work on the grid points chosen. A more analytic description for the the generalization of the matrix method is given in Appendix \ref{App:generalized_matrix_method}.
	
	\begin{figure}[t]\vspace{0.3cm}
		\includegraphics[width=0.5\textwidth]{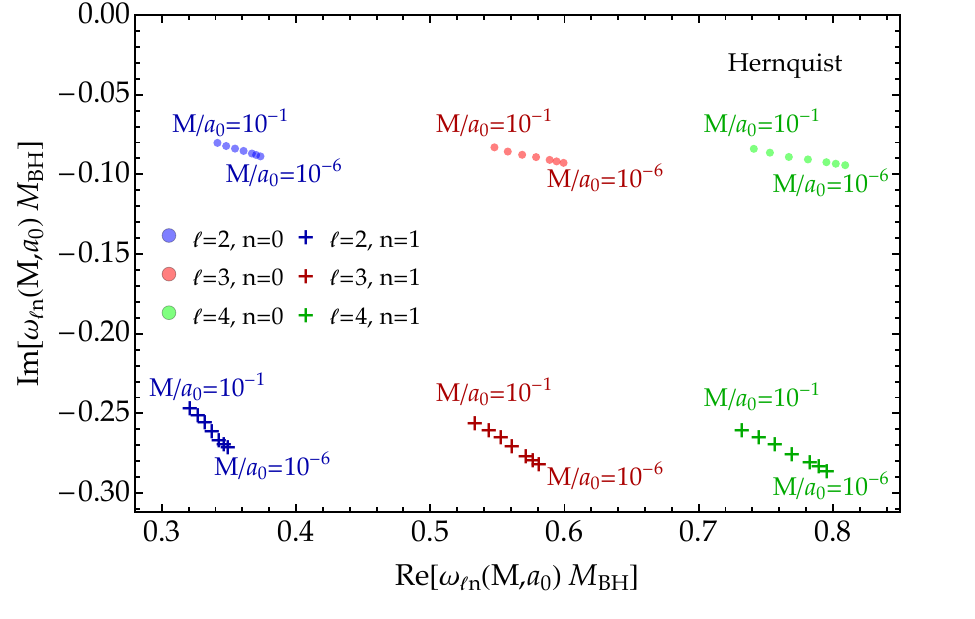}
		\caption{Axial QNMs $\omega_{\ell n}(M,a_0)$ with varying $\ell$ and overtone number $n$ of a numerical Hernquist-type BH, with $(\alpha,\beta,\gamma)=(1,4,1)$, as a function of the halo compactness $M/a_0$ computed with the generalized matrix method. Different colors correspond to different $\ell$, while points (crosses) correspond to the fundamental mode, $n=0$ (first overtone, $n=1$). The compactness shown are $M/a_0=(10^{-6}, 10^{-2}, 2 \times 10^{-2}, 4 \times 10^{-2}, 6 \times 10^{-2}, 8 \times 10^{-2}, 10^{-1})$.
		}\label{fig:Num_Hernquist_QNMs}
	\end{figure}
	
	Using the compactified radial coordinate $x$ described in the previous section, and applying the boundary conditions in Eq. (\ref{eq:boundNum}) by multiplying with the correct asymptotic behavior of the solutions of the master equation (\ref{eq:master}), one ends up with a homogeneous differential equation
	\begin{equation}
		\tau_0(x) \phi''(x) + \lambda_0(x) \phi'(x) + \sigma_0 (x) \phi(x)=0,
		\label{eq:MatrixEq}
	\end{equation}
	where $\tau_0(x),\, \lambda_0(x),\, \sigma_0 (x)$ are $\omega$-dependent functions and $\phi(x)$ is a general radial function. By discretizing the interval $x\in [0,1]$, it is possible to introduce $N$ random or evenly 
	distributed points with $x_1=0$ and $x_N=1$. Cramer's rule allows to discretize all derivatives \cite{Lin:2016,Lin:2016sch} and rewrite Eq. (\ref{eq:MatrixEq}) in matrix form, as $\mathcal{M} \mathcal{F}=0$, where $\mathcal{M}$ is a square matrix that depends on the eigenvalues $\omega$ of the system and $\mathcal{F}=\left(f_1,f_2,f_3, \dots ,f_N \right)^\mathrm{T}$ with $f_i=\phi(x_i)$. The homogeneous matrix equation should satisfy $\det(\mathcal{M})=0$. The latter provides an algebraic equation to calculate the eigenvalues of any particular eigenvalue system, which can be solved numerically. The generalized boundary conditions and the generic functions 
	$\tau_0(x), \lambda_0(x), \sigma_0 (x)$ for 
	asymptotically-flat, numerically-constructed BH 
	metrics are shown in Appendix \ref{App:generalized_matrix_method}.
	
	\begin{figure}[t]
		\includegraphics[width=0.5\textwidth]{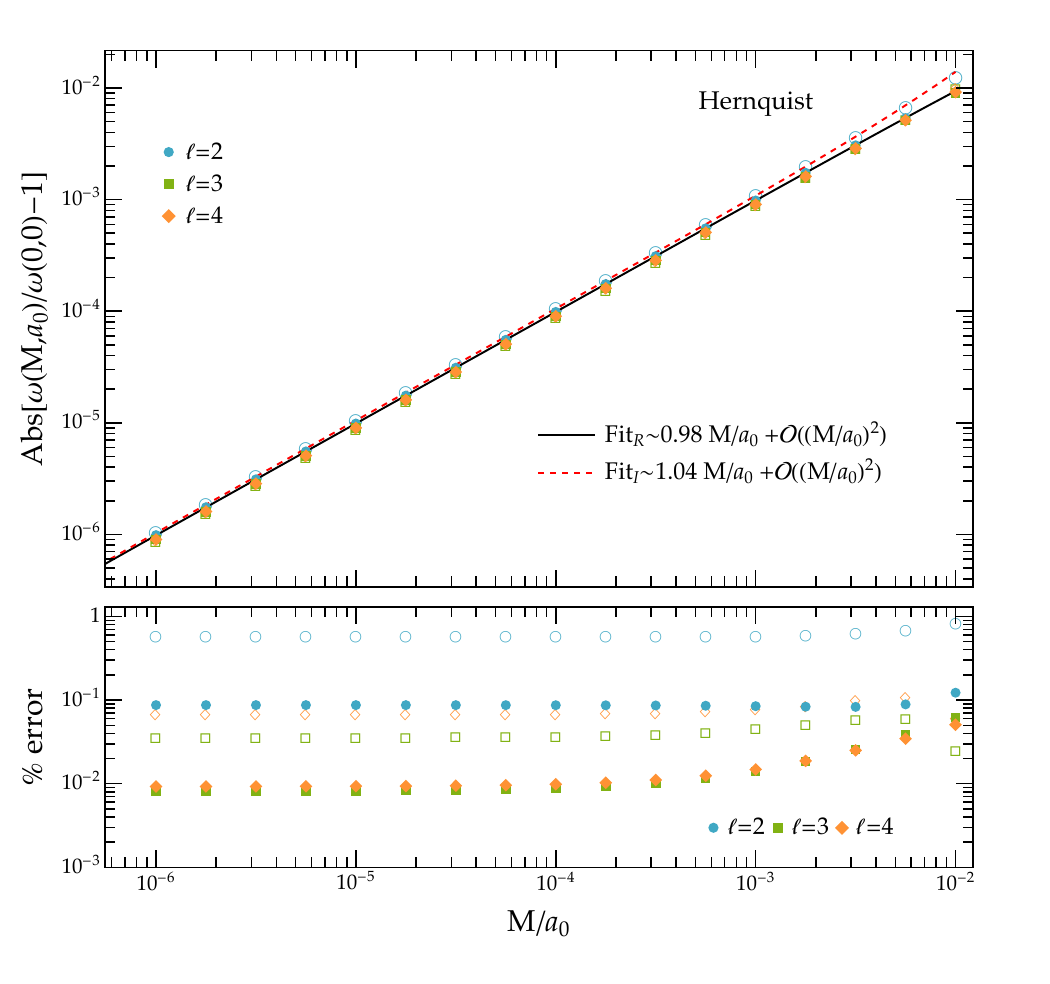}
		\caption{\emph{Top panel:} Relative difference between the real part (filled dots) and the imaginary part (empty dots) of the $n=0$ fundamental axial QNMs computed for Hernquist-type, with $(\alpha,\beta,\gamma)=(1,4,1)$, and vacuum 
			Schwarzschild BHs. We assume halo compactness between $M/a_0=10^{-6}$ and $M/a_0=10^{-2}$. Solid (and dashed) lines correspond to linear, semi-analytic fits of our data, with coefficients shown in the legend.\emph{Bottom panel:} Relative percentage difference between the $n=0$ Hernquist-type BH QNM frequencies 
			computed with the generalized matrix method and the 6th order WKB approximation, as a function of the compactness.}
		\label{fig:HernNum}
	\end{figure}
	
	Hereafter, we compute axial QNMs using $N=19$ 
	discretization points, keeping $70$ digits of precision in all internal calculations. Convergence tests have shown that QNM frequencies converge to a given value as we increase the number of grid points, as shown in Fig.~\ref{fig:Convergence_Matrix_all} of Appendix \ref{App:convergence}. We also compare numerical results obtained with the generalized matrix method with those obtained through the 6th Wentzel-Kramers-Brillouin (WKB) method. The WKB reproduces \textit{exact} QNM frequencies with good accuracy in the eikonal limit, i.e. for large $\ell$. The percentage error between axial QNMs from the generalized matrix method and WKB is always, at least, less than $1\%$.
	
	\section{Results}
	\subsection{A result of sufficient generality concerning light ring \& QNM frequencies in the eikonal regime}
	Before discussing QNMs, we will show that for {\it any} sufficiently dilute halo of matter, the light ring frequency and Lyapunov timescale are redshifted, by an amount that scales with the halo compactness. Since light ring properties are intimately connected to QNMs~\cite{Cardoso:2008bp}, this result then provides a robust basis to expect that to leading order QNM frequencies will be redshifted by environments.
	
	\begin{figure}[t]\vspace{0.3cm}
		\includegraphics[width=0.5\textwidth]{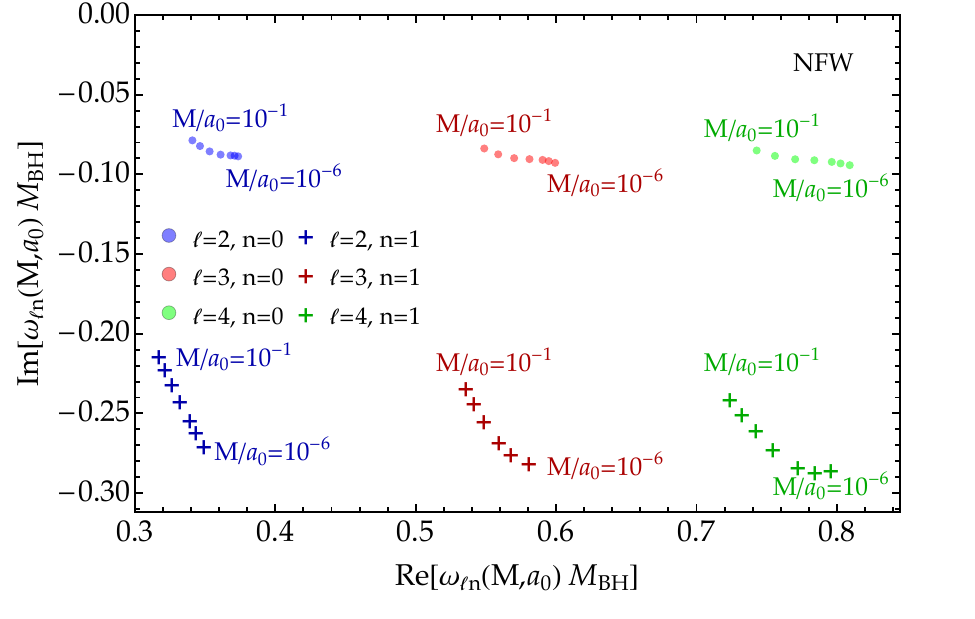}
		\caption{Axial QNMs $\omega_{\ell n}(M,a_0)$ with varying $\ell$ and overtone number $n$ of a numerical NFW-type BH with $(\alpha,\beta,\gamma)=(1,3,1)$, as a function of the halo compactness $M/a_0$ estimated with the generalized matrix method. Different colors correspond to different $\ell$, while points (crosses) correspond to the fundamental mode, $n=0$ (first overtone, $n=1$). The compactness shown are $M/a_0=(10^{-6}, 10^{-2}, 2 \times 10^{-2}, 4 \times 10^{-2}, 6 \times 10^{-2}, 8 \times 10^{-2}, 10^{-1})$.}
		\label{fig:Num_NFW_QNMs}
	\end{figure}
	
	Consider the general weak-field expression~\eqref{weak_field_geometry}. 
	The light ring location is determined by the condition $2f-rf'=0$, with a prime standing for a radial derivative. Using~\eqref{weak_field_geometry} we find to dominant order in the Newtonian potential that $r=3M_{\rm BH}(1+MU_{\rm LR}')$ with $U'=U'(r=3M_{\rm BH})$, leading to the orbital frequency (as measured by distant observers),
	\be
	M_{\rm BH}\Omega_{\rm LR}=\frac{1+U_{\rm LR}}{3\sqrt{3}},\label{eq:frequency_LR}
	\ee
	and a Lyapunov scale $\lambda$
	\be
	M_{\rm BH}\lambda=\frac{1+U_{\rm LR}}{3\sqrt{3}}.\label{eq:timescale_LR}
	\ee
	Notice that for a mass $M$ distributed on a scale $a_0\gg M_{\rm BH}$ then $U_{\rm LR} \sim U(0)\sim -M/a_0$, and indeed this coincides with the large $a_0$ limit of the exact solution \eqref{eq:f_Cardoso}. The potential $U_{\rm LR}$ gives the redshift of a photon as it crosses the galaxy, due to its matter content. The above -- a key result of this work which explains our numerics below -- then indicates that the leading order consequence of matter distributions is a redshift relative to vacuum modes, as had also been observed via time-domain signals~\cite{Spieksma:2024voy}. 
	
	\subsection{Axial QNMs of BHs within Hernquist and NFW matter profiles} \label{Hernquist}
	
	\begin{figure}[t]
		\includegraphics[width=0.54\textwidth]{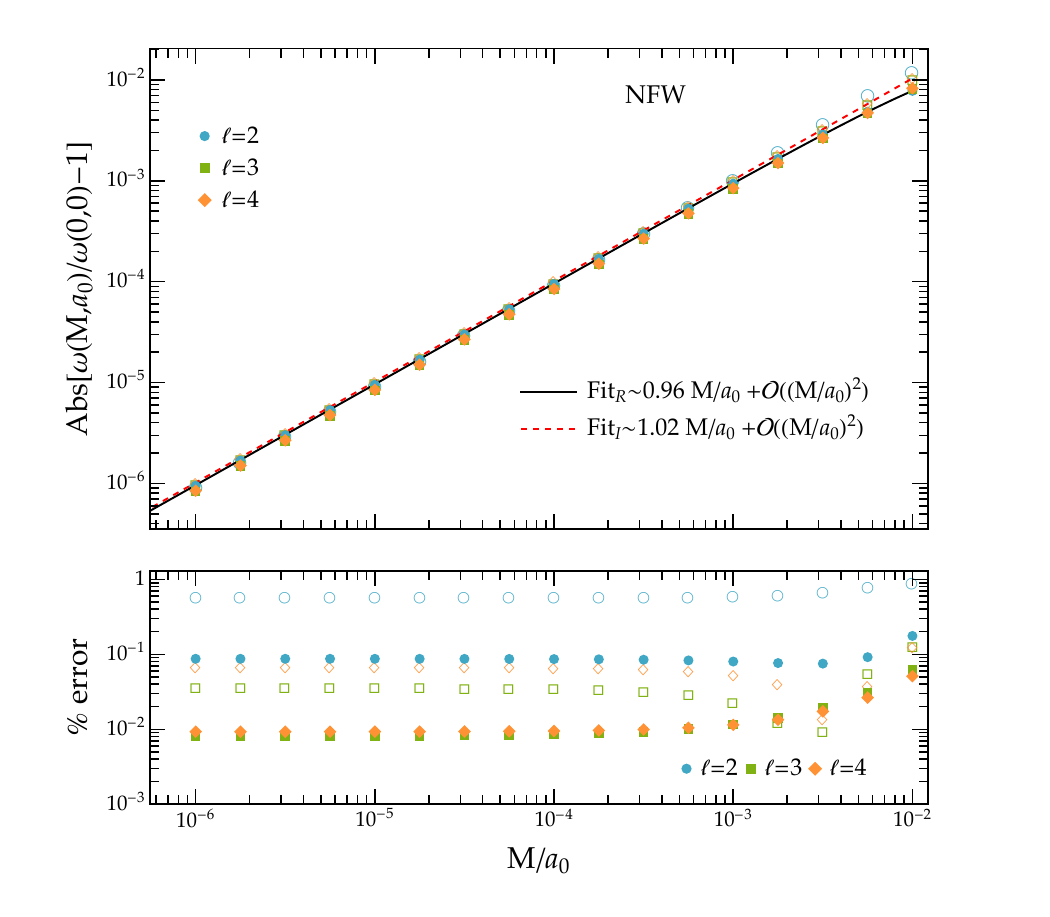}
		\caption{Same as Fig.~\ref{fig:HernNum} but for 
			NFW-type BH solutions with $(\alpha,\beta,\gamma)=(1,3,1)$.}    \label{fig:NFWNum}
	\end{figure}
	
	Our main numerical results are summarized in Figs.~\ref{fig:Num_Hernquist_QNMs}--\ref{fig:NFWNum}. Figure~\ref{fig:Num_Hernquist_QNMs} shows the real and the imaginary part of the fundamental and first overtone for Hernquist-type BHs, with varying $\ell$ and halo compactness $M/a_0$, where $(\alpha,\beta,\gamma)=(1,4,1)$. As expected, increasing $\ell$ yields QNM frequencies with larger real component, approaching the angular frequency of null particles at the light ring \cite{Cardoso:2008bp}. Our results show that more compact halo configurations lead to QNMs with lower (in absolute value) real and imaginary part, regardless of the overtone number and of the mode index $\ell$.

	The upper panel of Fig.~\ref{fig:HernNum} shows the relative difference of the QNM frequencies with respect to vacuum, for BHs of same mass $M_{\rm BH}$ for the Hernquist-type and profiles, as a function of $M/a_0$. The trend in Fig.~\ref{fig:HernNum} is consistent with a 
	\textit{redshift} scaling of both the real and the imaginary part of the modes according to the following linear expansion:
	\begin{equation}
		\frac{\omega_{\ell n}(M,a_0)}{\omega_{\ell n}(0,0)}\sim 1-\frac{M}{a_0}+\mathcal{O}\left(\frac{M}{a_0}\right)^2.\label{eq:redshift}
	\end{equation}
	The accuracy of the relation~\eqref{eq:redshift} 
	improves for realistic environments with $M/a_0\leq 10^{-4}$. This confirms results obtained for the exact BH metric \eqref{eq:uncharged_metric}, for which QNMs have been computed using a hyperboloidal foliation scheme~\cite{Ansorg:2016ztf,PanossoMacedo:2018hab,PanossoMacedo:2019npm,Hennig:2020rns,Jaramillo:2020tuu,PanossoMacedo:2024nkw} and a spectral collocation method~\cite{Jansen:2017oag,Cardoso:2017soq,Cardoso:2018nvb,Destounis:2018qnb,Liu:2019lon,Destounis:2019hca,Destounis:2019omd}, proving the accuracy of our generalized matrix method \cite{Cardoso:2008bp,Ringdown_review}. The lower panel of Fig.~\ref{fig:HernNum} also shows the relative difference between frequencies computed with the WKB approach, which differ, at most, less than $1\%$ from the generalized matrix method values.
	
	To further clarify the redshift effect on axial QNMs we consider an environment with $M/a_0\ll 1$ and study the master equation \eqref{eq:master} in the small compactness regime. To linear order in $M/a_0$, the 
	tortoise coordinate reads  $dr/dr_*\sim(1- M/a_0)\,dr/dr^\textrm{vac}_*$, where $r^\textrm{vac}_*$ is the tortoise coordinate of a vacuum Schwarzschild BH. Then, by expanding the scattering equation in $M/a_0$ we obtain:
	\begin{equation}
		\frac{d^2\psi_{\ell m}(r)}{d(r^\textrm{vac}_*)^2}+\left(\frac{\omega^2}{\gamma^2}-V_\textrm{ax}^\textrm{Schw}\right)\psi_{\ell m}(r)=0,
	\end{equation}
	where we have introduced the redshift factor 
	$\gamma\equiv 1-M/a_0$ and \cite{Regge:1957td}:
	\begin{equation}
		V_\textrm{ax}^\textrm{Schw}=\left(1-\frac{2M_\textrm{BH}}{r}\right)\left(\frac{\ell(\ell+1)}{r^2}-\frac{6M_\textrm{BH}}{r^3}\right).
	\end{equation}
	Therefore, to linear order in $\gamma$, the axial ringdown signal from a BH surrounded by a realistic environment is identical to that from a Schwarzschild BH, with redshifted frequency and mass, such that $\omega^\textrm{vac}\rightarrow\omega/\gamma$, that is perfectly consistent with the relation \eqref{eq:redshift} to first order.
	
	The axial QNMs for NFW-type BHs, with $(\alpha,\beta,\gamma)=(1,3,1)$, are drawn in 
	Fig. \ref{fig:Num_NFW_QNMs}. Their behavior in the complex plane as a function of $\ell$ and of the halo compactness $M/a_0$ is similar to that of Hernquist-type QNMs. More intriguingly QNMs exhibits the same redshift effect. Figure~\ref{fig:NFWNum} shows the relative 
	change $\omega_{\ell n}(M,a_0)/\omega_{\ell n}(0,0)$ as a function of $M/a_0$ for NFW-type BHs. The linear order scaling \eqref{eq:redshift} captures the deviation from the vacuum case across a wide range of compactness, such that $\omega^\textrm{vac}\rightarrow \omega/\gamma$. This suggests that the redshift effect is a generic phenomenon affecting the axial QNMs of BHs surrounded by a spherically symmetric matter distribution. Indeed, the generic result \eqref{eq:frequency_LR}--\eqref{eq:timescale_LR} for the properties of the light ring also describes the QNM frequencies in the eikonal regime, and we find it is a very good description of even the fundamental mode.
	
	\section{Conclusions}
	``Listening'' to the Universe with QNMs is the foundation of BH spectroscopy. Decades of developments in the field of ringdown physics has provided significant information regarding the ``death'' of stars and the formation of BHs through binary coalescence. In the near future, we expect that the BH spectroscopy program will be the most undisputed scheme for analyzing the relaxation of perturbed compact objects and understanding the nature of binary remnants. Nevertheless, only recently BH spectroscopy started taking into account the effects of surrounding astrophysical environments and galactic matter when calculating QNMs.
	
	In fact, since almost all BHs reside in some sort of environment (accretion disks, galactic and extra-galactic) the effort towards a complete BH spectroscopy program that includes environmental effects is undergoing rapid advancements. Very recently, a fully-relativistic, static and spherically symmetric BH was found that is embedded in a spherical matter halo \cite{Cardoso:2021wlq}. This solution does not contain \emph{impromptu} post-Newtonian additions to the metric that mimic environmental effects. To the contrary, it is a fully-relativistic BH solution of the Einstein field equations surrounded by an anisotropic generic fluid.
	
	Understanding the environmental effect on BH QNMs is now very timely and of utmost importance for current and future detections of GWs. In fact, a spurt of analyses have taken place very recently for a unique, exact, BH solution surrounded by a Hernquist-type cold matter \cite{Cardoso:2021wlq,Konoplya:2021ube,Cardoso:2022whc,Cardoso:2024mrw}. Since this solution is the only one that currently exist in closed form (besides some thin-shell matching solutions \cite{Jusufi:2022jxu} and analytic approximations \cite{Konoplya:2022hbl}), using different matter galactic halos leads to numerical BH solutions surrounded by these matter halos \cite{Figueiredo:2023gas,Speeney:2024mas}.
	
	Here, we have initiated the use of numerically-constructed BH configurations with different matter halos to find their \emph{axial QNMs} and observe the overall effect of the environment to the modes. We show that BHs with environments possess scalar, electromagnetic and axial gravitational QNMs that are governed by a single-parameter effective potential. The analyses in \cite{Cardoso:2021wlq,Ringdown_review} have shown that the exact solution from Ref. \cite{Cardoso:2021wlq} possesses scalar and axial QNMs that are \emph{redshifted}, both in frequency and (absolute value of) decay timescale. In this work, we solidify that this effect on axial QNMs is universal, no matter the energy density (in our study, numerical Hernquist and numerical NFW-type BHs). This is shown by the use of matrix method from Ref. \cite{Lin:2016,Lin:2016sch} which we generalized so that it works for numerically-constructed backgrounds. The code is publicly available in Ref. \cite{git_axial} and can be used not only for environmental effects, but also for other kinds of solutions such as hairy BHs that do not possess an exact, closed form, and BHs with accretion disks.
	
	Overall, our code behaves very well, in accord to all tests we have performed, and produces convergent QNMs (see Appendix \ref{App:convergence}). Since axial perturbations do not couple to the surrounding matter, a redshift of modes, as compactness increases, is adequate. Therefore, the difference between axial QNMs due to environmental effects can be equivalent to redshifted vacuum Schwarzschild QNMs. However, we expect redshift scaling to break for polar perturbations, which couple to matter variables \cite{Speeney:2024mas}. The polar sector of gravitational perturbations couples to matter and this will enable us to test GR in its strongest regime. Calculations of polar QNMs for different families of matter profiles are underway and will be addressed in a followup publication, along with an assessment 
	of their detectability by future GW observations. 
	
	\emph{Note:} During the completion of our work, Ref. \cite{Chakraborty:2024gcr} appeared, that contains results for axial QNMs, for an exact metric, that are partially comparable to ours.
	
	\begin{acknowledgements}
		K.D. acknowledges partial support by the MUR PRIN Grant 2020KR4KN2 “String Theory as a bridge between Gauge Theories and Quantum Gravity” and by the MUR FARE programme (GW-NEXT, CUP: B84I20000100001). K.D. also acknowledges financial support provided by FCT – Fundação para a Ciência e a Tecnologia, I.P., under the Scientific Employment Stimulus – Individual Call – Grant No. 2023.07417.CEECIND. A.M. acknowledges financial support from MUR PRIN Grant No. 2022-Z9X4XS, funded by the European Union - Next Generation EU. We acknowledge support by VILLUM Foundation (grant no. VIL37766) and the DNRF Chair program (grant no. DNRF162) by the Danish National Research Foundation. V.C.\ is a Villum Investigator and a DNRF Chair. V.C. acknowledges financial support provided under the European Union’s H2020 ERC Advanced Grant “Black holes: gravitational engines of discovery” grant agreement no. Gravitas–101052587. Views and opinions expressed are however those of the author only and do not necessarily reflect those of the European Union or the European Research Council. Neither the European Union nor the granting authority can be held responsible for them. This project has received funding from the European Union's Horizon 2020 research and innovation programme under the Marie Sklodowska-Curie grant agreement No 101007855 and No 101131233.
	\end{acknowledgements}
	
	\begin{appendices} 
		
		\section{Generalized matrix method} \label{App:generalized_matrix_method}
		
		One of the numerical computation techniques to solve eigenvalue problems of differential equations is the matrix method \cite{Lin:2016,Lin:2016sch}. With this method, it is possible to transform the Schrödinger-like equation into a matrix equation via the proper manipulation of the ordinary differential equation that governs perturbation propagation through proper decomposition of the spatial derivatives. 
		
		The proposed matrix method in \cite{Lin:2016,Lin:2016sch} is a non-grid-based interpolation approach, where the interpolation is based on Taylor expansions. This method discretizes the spatial coordinate so that the differential equation, as well as its boundary conditions, are transformed into a homogeneous matrix equation. With this procedure the eigenfunction is expanded in the vicinity of all grid points, and therefore the precision of the algorithm is potentially improved. The data points do not need to sit on a specified grid for the method to work (equispaced in \cite{Lin:2016sch}, delimited expansion \cite{Lin:2022ynv} and Chebyshev points \cite{Shen:2022ssv}). Using the information of a set of $N$ randomly scattered data points, Taylor series are carried out for the unknown eigenfunction up to $N$-th order for each data point. Then, the resulting homogeneous system of linear algebraic equations is solved for the eigenvalues. 
		
		The main advantage of this method is that, choosing the right coordinate transformations, the discretization of the wave function and its derivatives are made to be independent of any specific metric, as long as we are aware of the spacetime's asymptotic nature. Since our goal is to carry out an interpolation based on the information of a set of $N$ scattered points distributed in a small neighborhood, the Taylor series has to be applied $N$ times to each one of the data points around $x_0$. The resulting equation can be written as a matrix product
		\begin{equation}
			\mathcal{F}=\mathcal{M} D,
		\end{equation}
		where $\mathcal{F}$ and $D$ are $N \times 1$ column vectors and $\mathcal{M}$ is a $N \times N$ matrix. $\mathcal{F}$ contains the value of the function $f(x)$ at each data point $N$, $D$ contains the value of $f(x)$ and its derivatives at $x_0$ and the matrix $\mathcal{M}$ consists of $N$ rows such that the $i$-th row contains increasing power functions of the coordinate relative difference between the $i$-th data point and the query point $x_0$. 
		
		If one considers a univariate function $f(x)$ and the data points as function values at coordinates $x_i$, with $i=1,2, \dots, N$, then the vectors $\mathcal{F}$, $D$ and the matrix $\mathcal{M}$ can be written as
		\begin{equation}
			\mathcal{F}=\left( f(x_1), f(x_2), f(x_3), \dots , f(x_N) \right)^\mathrm{T},
		\end{equation}
		\begin{equation}
			\mathcal{M}=\begin{pmatrix}
				1 & x_1-x_0 & \frac{(x_1-x_0)^2}{2} & \frac{(x_1-x_0)^3}{3!}  & \dots & \frac{(x_1-x_0)^N}{N!} \\
				1 & x_2-x_0 & \frac{(x_2-x_0)^2}{2} & \frac{(x_2-x_0)^3}{3!} & \dots & \frac{(x_2-x_0)^N}{N!} \\
				1 & x_3-x_0 & \frac{(x_3-x_0)^2}{2} & \frac{(x_3-x_0)^3}{3!} & \dots & \frac{(x_3-x_0)^N}{N!} \\
				\vdots & \vdots & \vdots & \vdots & \ddots & \vdots \\
				1 & x_N-x_0 & \frac{(x_N-x_0)^2}{2} & \frac{(x_N-x_0)^3}{3!} & \dots & \frac{(x_N-x_0)^N}{N!} 
			\end{pmatrix},
		\end{equation}
		\begin{equation}
			D=\left( f(x_0), f'(x_0), f''(x_0), \dots , f^{(N)}(x_0) \right)^\mathrm{T}.
		\end{equation}
		If the determinant of $\mathcal{M}$ is non-zero, the column vector $D$ can be expressed in terms of $\mathcal{F}$ and $\mathcal{M}$ as $D=\mathcal{M}^{-1}\mathcal{F}$. In particular, if one is interested in only a few particular elements (derivatives) from $D$, it is possible to use Cramer's rule to evaluate them
		\begin{equation}
			D_i=\frac{\det(\mathcal{M}_i)}{\det(\mathcal{M})},
			\label{eq:Cramer}
		\end{equation}
		where $\mathcal{M}_i$ is the matrix formed by replacing the $i$-th column of $\mathcal{M}$ by the column vector $\mathcal{F}$. For instance, $f'(x_0)=\det(\mathcal{M}_2)/\det(\mathcal{M})$.
		This way, it is possible to rewrite the $n$-th order derivatives $f^{(n)}(x)$ using (\ref{eq:Cramer}) and all the derivatives of the $N$ data points are written as linear combinations of the function values $f(x_i)$. Substituting the derivatives into the eigenequation in study, one obtains $N$ equations with $f(x_1), f(x_2), \dots , f(x_N)$ as its variables.
		
		After properly applying the boundary conditions by multiplying with the correct asymptotic behavior of the solutions of the master equation (\ref{eq:master}), one ends up with a homogeneous differential equation
		\begin{equation}
			\tau_0(x) \phi''(x) + \lambda_0(x) \phi'(x) + \sigma_0 (x) \phi(x)=0.
		\end{equation}
		When dealing with numerically-constructed backgrounds of static and spherically-symmetric BHs embedded in generic matter halos, one first obtains the data points of all metric tensor components, as discussed in Sec. \ref{numerical_method}, interpolate them so that they can be used as functions in the rest of the calculations. and tweak the internal precision of every calculation to $70$ digits for increased accuracy. These interpolation functions can then be used as normal functions at any grid point we desire. Then, by taking into account incoming waves at the event horizon and outgoing waves at infinity, we find the form that solutions must satisfy at the boundaries in the wave function
		\begin{equation}
			\phi(x)=e^{-\frac{i r_\textup{h} \omega }{x-1}} (1-x)^{-i \omega  (2 M+r_\textup{h} )} x^{-\frac{i r_\textup{h}  \omega }{\sqrt{r_\textup{h}  f'(r_\textup{h})}}} R(x).
			\label{eq:wave_num}
		\end{equation}
		In this case, the boundary conditions become $R(0) = R_0$ and $R(1) = R_1$, where $R_0$ and $R_1$ are indeterminate constants. In accordance with the boundary conditions, we can rewrite the wave function as
		\begin{equation}
			\chi(x)=x (1-x) R(x).
			\label{eq:simplification}
		\end{equation}
		This leads to a further simplification, since $\chi(0)=\chi(1)=0$. The new boundary conditions guarantees that the resulting matrix equation is homogeneous. The final master equation can be expressed as
		\begin{equation}
			\tau(x)\chi''(x)+\lambda(x)\chi'(x)+\sigma(x)\chi(x)=0.
		\end{equation}
		
		For the static and spherically-symmetric backgrounds we are considering, the matrix coefficients have the generic form
		\begin{widetext}
			\begin{align*}
				\tau(x)=& f\left(\frac{-r_\textup{h}}{-1 + x}\right) \left[2 m\left(\frac{-r_\textup{h}}{-1 + x}\right) + \frac{r_\textup{h}}{-1 + x}\right] (1 - x)^5 x,
				\\
				\lambda(x)=& \frac{1}{\sqrt{r_\textup{h} f'\left(r_\textup{h}\right)}}f\left(\frac{-r_\textup{h}}{-1 + x}\right)(1 - x)^2\bigg[2 i r_\textup{h}^2 (-1 + x)^2 \omega + 2 (-1 + x) \left(\frac{-r_\textup{h}}{-1 + x}\right) \bigg((-1 + x) (-2 + x + 4 i M x \omega)\sqrt{r_\textup{h} f'(r_\textup{h})}\\
				&+ 2 i r_\textup{h} \omega \left(1 + (-2 + x) x (1 + \sqrt{r_\textup{h} f'(r_\textup{h})}\,)\right)\bigg) + r_\textup{h} \sqrt{r_\textup{h} f'(r_\textup{h})} \bigg(2 + 2 x (-2 + x + i (r_\textup{h} (-2 + x) + 2 M (-1 + x)) \omega)\\
				&- (-1 + x) x \, m'\left(\frac{-r_\textup{h}}{-1 + x}\right)\bigg)\bigg],
				\\
				\sigma(x)=& \frac{1}{x f'(r_\textup{h})}f\left(\frac{-r_\textup{h}}{-1+x}\right) \left(r_\textup{h}+ 2(-1+x)m\left(\frac{-r_\textup{h}}{-1+x}\right)\right)\bigg[\bigg(-2 + 2 x (5 + 2 i M \omega + 2 i r_\textup{h}\omega) - 2 x^3 (-9 - 9 i r_\textup{h} \omega + 4 M^2 \omega^2\\
				&+ 2 r_\textup{h}^2 \omega^2 + 6 M \omega (-2 i + r_\textup{h} \omega))+ x^4 \bigg(-6 - 5 i r_\textup{h} \omega + 4 M^2 \omega^2 + r_\textup{h}^2 \omega^2 + 2 M \omega (-5 i + 2 r_\textup{h} \omega)\bigg) + x^2 \bigg(-20 - 17 i r_\textup{h} \omega + 4 M^2 \omega^2\\
				&+ 4 r_\textup{h}^2 \omega^2+ 2 M \omega (-9 i + 4 r_\textup{h} \omega)\bigg)\bigg) f'(r_\textup{h}) + (-1 + x)^2 \omega \bigg((-1 + x) (3 i + x (-5 i + 4 M \omega)\bigg) \sqrt{r_\textup{h} f'(r_\textup{h})} + r_\textup{h} \omega \bigg(1\\
				&- (2 x- x^2) (1 + 2 \sqrt{r_\textup{h}f'(r_\textup{h})})\bigg)\bigg)\bigg] + \frac{1}{\sqrt{r_\textup{h}f'(r_\textup{h})}}\bigg[(-1 + x) f\left(\frac{-r_\textup{h}}{-1 + x}\right) ((-1 + x) (-1 + x (2 + 2 i M \omega)) \sqrt{r_\textup{h}f'(r_\textup{h})}\\
				&+ i r_\textup{h} \omega (1 - (2 x -x^2) (1 + \sqrt{r_\textup{h}f'(r_\textup{h})}))) (6 (-1 + x) m\left(\frac{-r_\textup{h}}{-1 + x}\right)+ r_\textup{h} (2 + m'\left(\frac{-r_\textup{h}}{-1 + x}\right)))\bigg]\\
				&+(1 - x)^3 x \left[\frac{r_\textup{h}^3 \omega^2}{(-1 + x)^3} + f\left(\frac{-r_\textup{h}}{-1 + x}\right)\left(-6 m\left(\frac{-r_\textup{h}}{-1 + x}\right) - \frac{r_\textup{h} \left(l + l^2 + m'\left(\frac{-r_\textup{h}}{-1 + x}\right)\right)}{-1 + x}\right)\right]. 
			\end{align*}
		\end{widetext}
		
		By using the Cramer's rule, it is possible to discretize all derivatives \cite{Lin:2016,Lin:2016sch} and rewrite (\ref{eq:MatrixEq}) in the matrix form
		\begin{equation}
			\mathcal{M} \mathcal{F}=0,
		\end{equation}
		where $\mathcal{M}$ is a square matrix that depends on the eigenvalues $\omega$ of the system (in our case the QNMs), while $\mathcal{F}=\left(f_1,f_2,f_3, \dots ,f_N \right)^\mathrm{T}$ with $f_i=\phi(x_i)$. In order to have non-trivial solutions, this homogeneous matrix equation should satisfy
		\begin{equation}
			\det(\mathcal{M})=0.
			\label{eq:EigenMatrix}
		\end{equation}
		The matrix equation \eqref{eq:EigenMatrix} leads to and algebraic equation that depends on powers of $\omega$. By solving it we obtain the BH QNMs.
		
		\section{QNM convergence}\label{App:convergence}
		
		Here, we perform the following convergence tests: (i) we find the axial QNMs of the exact  BH from Eq. \eqref{eq:uncharged_metric} with the standard matrix method and compare them with the those obtained with the generalized matrix method for Hernquist-type BHs, and (ii) we perform convergence tests of the fundamental modes of Hernquist-type and NFW-type BHs by increasing the number of grid points $N$.
		
		\begin{figure}[t]
			\includegraphics[width=0.46\textwidth]{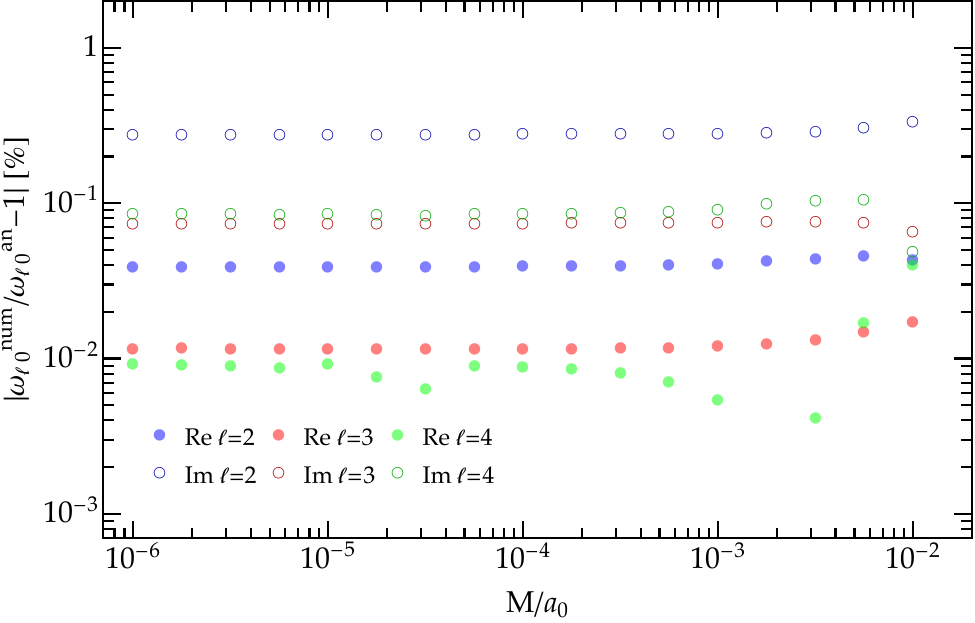}
			\caption{Percentage error between the real (filled circles), and imaginary (empty circles) part, of Hernquist axial $n=0$ fundamental QNMs calculated with the standard matrix method by using the analytical metric in Eq. \eqref{eq:uncharged_metric}, $\omega_{\ell n}^\text{an}$, and the generalized matrix method by using numerically-constructed, Hernquist-type metric, $\omega_{\ell n}^\text{num}$, for varying $\ell$, as a function of the halo compactness $M/a_0$. The compactness is varied from $M/a_0=10^{-6}$ to $10^{-2}$.} \label{fig:Convergence_QNMs}
		\end{figure}
		
		The convergence analysis (i) is performed by 
		computing QNMs for the exact spacetime 
		discussed in Sec.~\ref{sec:DManalytic} with 
		the standard matrix method, and QNMs from the 
		Hernquist-type BHs determined through 
		the procedure described in Sec. \ref{numerical_method}. The relative difference between the two sets of modes is shown in Fig. \ref{fig:Convergence_QNMs}. By comparing the fundamental modes, we find a percentage difference of order $\mathcal{O}(<1\%)$ for the real part and of $\mathcal{O}(<10^{-1}\%)$ for the imaginary part (by varying compactness $M/a_0$ from $10^{-6}$ to $10^{-2}$). We expect that equally small QNM percentage discrepancies will occur for different matter halos.
		
		\begin{figure}[t]
			\includegraphics[width=0.5\textwidth]{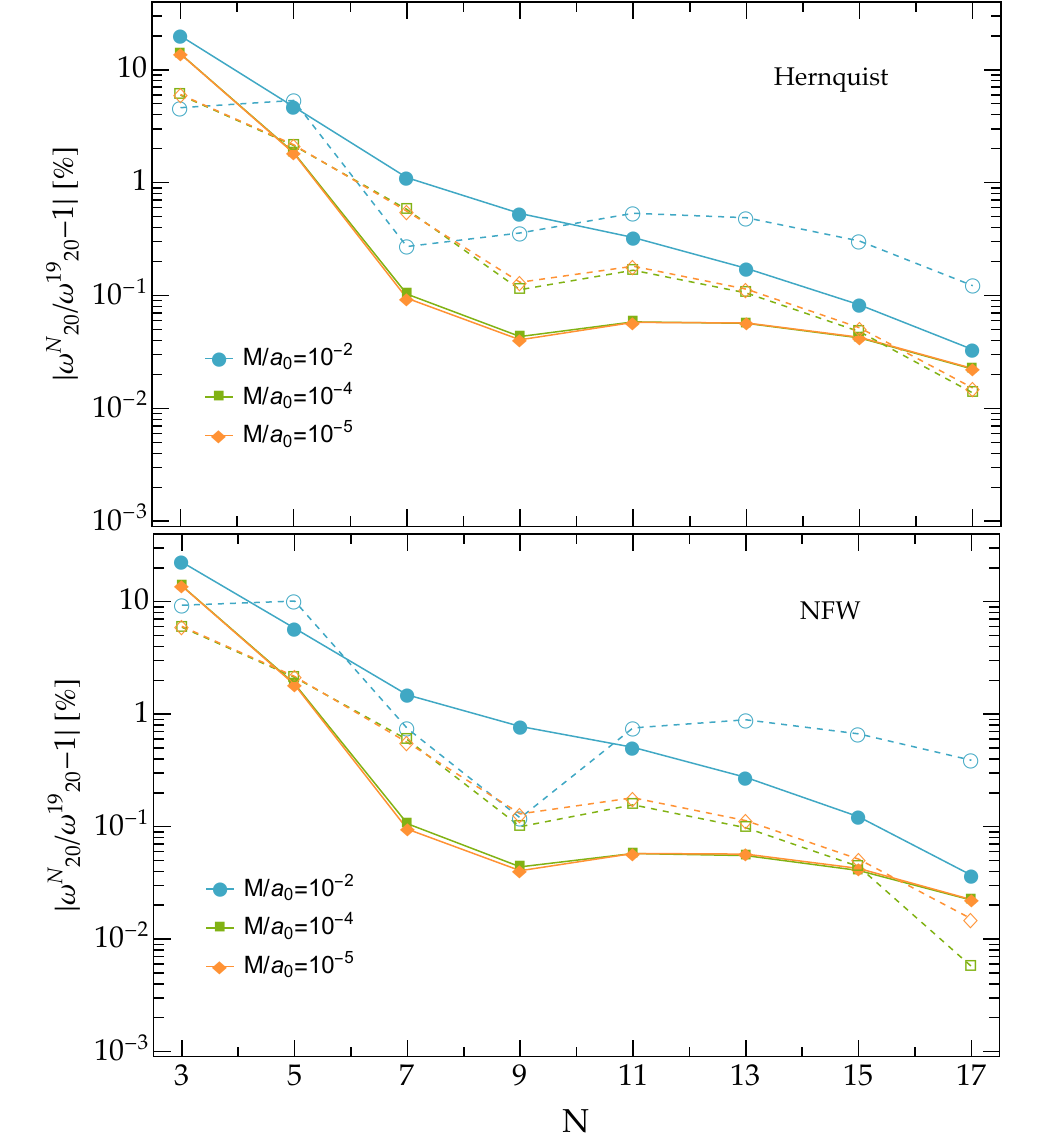}
			\caption{Convergence of $\ell=2$ fundamental QNM for the generalized matrix method when the number of grid points $N$ is increased in the range $[3,17]$ for Hernquist-type BHs (top panel), and NFW-type BHs (bottom panel). The fundamental modes are compared with the reference modes evaluated with $N=19$. The percentage difference is performed for three value for the compactness, i.e. $M/a_0=(10^{-5},\, 10^{-4},\,10^{-2})$.}\label{fig:Convergence_Matrix_all}
		\end{figure}
		
		Finally, Fig.~\ref{fig:Convergence_Matrix_all} 
		demonstrates the convergence of the fundamental 
		axial QNM for the Herquist and NFW-type BHs for different $M/a_0$. We assume that the modes resulting from the generalized matrix method with $N=19$ are accurate enough to be the reference modes and show that as $N$ increases then the axial $n=0$ QNMs indeed converge to those found and utilized throughout the paper. We observe that the percentage error gradually decreases as $N$ is increased from 3 to 17. The convergence is rather fast, especially when compared to other methods that use discretization techniques such as pseudospectral \cite{Jansen:2017oag} and hyperboloidal methods \cite{Ansorg:2016ztf}, where a grid point number of $N=50$ is needed to achieve the same accuracy as the one shown in Fig. \ref{fig:Convergence_Matrix_all}.
		
	\end{appendices}
	\bibliography{biblio}
	
\end{document}